\documentclass[twocolumn,showpacs,preprintnumbers,amsmath,amssymb,prb]{revtex4}
\usepackage{amsmath,amsfonts,bm,graphicx}
\usepackage{color}

\begin{document}

\title{Graphene antidot lattice waveguides}

\author{
Jesper Goor Pedersen$^1$, 
Tue Gunst$^{2,3}$,
Troels Markussen$^4$,
and Thomas Garm Pedersen$^{1,3}$
}

\affiliation{
$^1$
Department of Physics and Nanotechnology,
Aalborg University, Skjernvej 4A
DK-9220 Aalborg East, Denmark
\\
$^2$
Department of Micro- and Nanotechnology (DTU Nanotech),
Technical University of Denmark, 
DK-2800 Kgs. Lyngby, Denmark
\\
$^3$
Center for Nanostructured Graphene (CNG), 
Technical University of Denmark, 
DK-2800 Kgs. Lyngby, Denmark
\\
$^4$
Center for Atomic-scale Materials Design (CAMD), Department of Physics,
Technical University of Denmark, 
DK-2800 Kgs. Lyngby, Denmark
}

\date{\today}

\begin{abstract}
We introduce graphene antidot lattice waveguides: nanostructured graphene where
a region of pristine graphene is sandwiched between regions of graphene antidot lattices. The band gap
in the surrounding antidot lattices enable localized states to emerge in the central waveguide region.
We model the waveguides via a position-dependent mass term in the Dirac approximation of graphene, and arrive at
analytical results for the dispersion relation and spinor eigenstates of the localized waveguide modes.
To include atomistic details we also use a tight-binding model, which is in excellent agreement with the
analytical results. The waveguides resemble graphene nanoribbons, but without the particular properties of
ribbons that emerge due to the details of the edge. We show that electrons can be guided through kinks without 
additional resistance and that transport through the waveguides is robust against structural disorder.
\end{abstract}


\maketitle

\section{Introduction}
Graphene, the two-dimensional allotrope of carbon first isolated in 2004,\cite{Novoselov2004,Neto2009}
has emerged as a highly attractive material for future electronic devices.
Graphene has exceptional electronic properties, as demonstrated in its extremely high
carrier mobility,\cite{Morozov2008} which even at room temperature is limited predominantly 
by impurity scattering.\cite{Geim2007} Already, extremely fast graphene-based transistors have been
realized,\cite{Lin2010} and fabrication methods have emerged, which allow for large-scale production
of single-layered graphene.\cite{Bae2010}
One key element of future graphene-based electronics is the ability to localize
carriers in graphene \emph{wires}, in order to facilitate electronic graphene networks.
The most immediate way of realizing such wire structures is by cutting graphene into
so-called graphene nanoribbons (GNRs).\cite{Botello-Mendez2011} Quantum confinement will in general induce 
a band gap in such structures, the magnitude of which scales with the inverse of
the width of the GNR. However, the exact atomistic configuration of the
edge of the ribbon greatly influences the magnitude of this gap, with particular edge 
configurations resulting in vanishing gaps or localized edge states.\cite{Nakada1996,Brey2006}
An alternative to this is to rely on controlled generation of extended defects in graphene
to act as metallic wires.\cite{Lahiri2010}
Relying on results from two-dimensional electron gases (2DEGs) formed at the interface of semiconductor 
heterostructures, one could also imagine applying electrostatic gating to define wire geometries.
Graphene presents an interesting challenge in this regard via the phenomenon of Klein tunneling,\cite{Novoselov2006} which
makes it difficult to achieve carrier localization in graphene via ordinary gating.
Electrons impinging on a potential barrier at close to normal incidence are transmitted
with vanishing reflection, regardless of the height of the potential barrier.
In spite of this, guiding of electrons via electrostatic gating has been demonstrated
experimentally,\cite{Williams2011} albeit with the caveat that guiding is restricted
to a specific range of wave vectors for which Klein tunneling is negligible.
In contrast, a so-called \emph{mass term} in graphene provides confinement that is a close
analogue of gate-defined localization in an ordinary 2DEG.
This term originates from a Dirac description of graphene,
which emerges as a low-energy approximation of a tight-binding (TB) model on the honeycomb
lattice.\cite{Semenoff1984} Adding a diagonal term $\pm \Delta$ to the resulting Hamiltonian, 
with the sign  alternating between the two sublattices of graphene, the spectrum becomes that of
gapped graphene, a semi-conductor with an energy gap of twice the mass term. This 
staggered potential is commonly denoted a mass term due to the analogy of the low-energy carriers
in graphene with massless Dirac fermions, which acquire a mass via such a term in the
Hamiltonian.

\begin{figure}
\includegraphics[width=\linewidth]{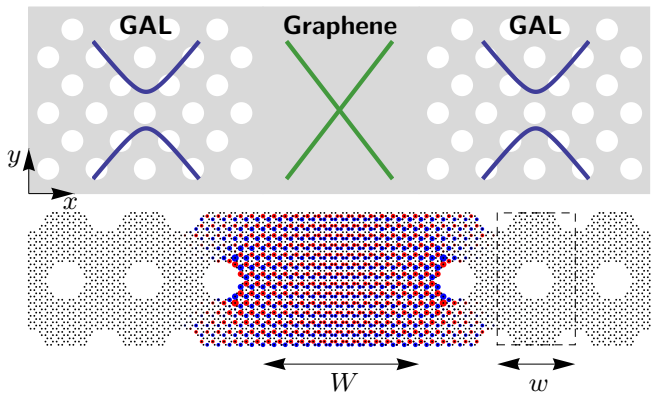}
\caption{
(upper panel) Conceptual illustration of a graphene antidot lattice (GAL) waveguide.
A central region of pristine graphene is surrounded by GAL regions, the band gaps of which confine
states to the waveguide region. Translational symmetry is assumed in the $y$--direction, which is along
the longitudinal direction of the waveguide.
(lower panel) The geometry of a $\{7,3\}_2^\mathrm{zz}$ GAL waveguide. Black dots show the location of carbon atoms.
Bloch boundary conditions are imposed on all boundaries.
The dashed lines illustrate the enlarged GAL unit cell, the width of which we denote by $w$.
The lowest-energy waveguide mode at the $\Gamma$ point, calculated via the tight-binding model, is illustrated with circles, the size of which shows the
absolute value $|\psi(x,y)|$ of the (real-valued) eigenstate, while the color indicates the sign.
}
\label{fig:concept}
\end{figure}
In this paper, we propose realizing a waveguiding structure in graphene via
graphene antidot lattices (GALs). GALs are nanostructured graphenes, which
in their simplest description take the form of periodically perforated graphene structures.\cite{Pedersen2008} 
GALs have recently been produced experimentally with both electron beam
lithography \cite{Eroms2009,Begliarbekov2011} and block copolymer lithography.\cite{Kim2010,Bai2010}
The periodic perforation induces a band gap in graphene, rendering it semi-conducting, and the
resulting band structure closely resembles that of gapped graphene in the low-energy
limit. GALs thus allow for the realization of a position-dependent mass term, as illustrated in
Fig.~\ref{fig:concept}. The idea is to sandwich a region of pristine graphene between
two GAL regions, the band gaps of which define an energy range for which localized guided modes
are expected to emerge in the central region. This idea is closely analogous to how photonic waveguiding 
is realized in photonic crystal structures.\cite{Chutinan2000}
We note that other methods beside GALs have been proposed for achieving gaps in graphene via superlattices, such as, e.g., 
patterned hydrogenation\cite{Balog2010} or superlattices of boron nitride islands embedded in graphene.\cite{Lopez2012}
Another alternative is to sandwich graphene between hexagonal boron nitride, which is predicted to induce
significant band gaps in graphene.\cite{Quhe2012}

We denote a waveguide geometry as $\{L,R\}_N^\mathrm{zz/ac}$, where $L$ is the sidelength of the hexagonal unit cell
of the surrounding GAL, while $R$ is the radius of the perforations, both in units of the graphene
lattice constant, $a=2.46$~\AA. The width $W$ of the waveguide is defined via $N\equiv W/w$, where 
$w$ is the width of the enlarged GAL unit cell, as illustrated in the lower panel of Fig.~\ref{fig:concept}.
The width of the waveguide is of course somewhat more ambiguous than in the case of GNRs, as we have no sharp
edge defining the precise boundary between the region of pristine graphene and the surrounding antidot lattice.
For denoting the geometries we simply take $W$ as the distance between the nearest edges of the two bordering
antidot lattice unit cells. However, as we discuss below, the effective waveguide width is slightly larger than this.
The width $w$ of the enlarged GAL unit cell depends on the orientation of the waveguide with respect to the graphene lattice, 
which we indicate with the superscript, with 'zz' ('ac') denoting a waveguide with the longitudinal direction along the 
zigzag (armchair) orientation of the graphene lattice.
Note that in both cases we choose the orientation of the GAL such that the
superlattice basis vectors lie parallel to carbon-carbon bonds, to ensure that a band gap always exists for the 
GAL.\cite{Petersen2011} The lower panel of Fig.~\ref{fig:concept} illustrates the
geometry of a $\{7,3\}_2^\mathrm{zz}$ waveguide. For simplicity we restrict the width of the waveguide to
be an integer multiple of the width of the GAL unit cell. This is merely for computational convenience, and
as we will demonstrate in this paper, simple scaling laws exist to predict the properties
of more general widths of the waveguide.

In this paper, two different methods will be employed to determine the waveguiding properties 
of GAL waveguides. We first consider a model based on the Dirac approximation, including the
influence of the GAL barriers via a position-dependent mass term. We will show that this model
admits analytical solutions in certain limits, which is highly beneficial for determining the general
dependence of the waveguide properties on the various structural parameters. Furthermore, these
results demonstrate clearly the unique properties of graphene waveguiding compared to quantum well
structures defined in ordinary 2DEGs. To include the atomistic details of the structures we also
consider a tight-binding model, which we use to calculate the transmission properties of the waveguides,
taking into account potential structural disorder.

\section{Dirac model}
\subsection{Analytical derivation}
We first consider a simple model of a GAL waveguide based on a Dirac model of graphene.
For graphene nanoribbons, the exact edge geometry can be included in a Dirac model via the boundary conditions 
of the spinor components in each valley.\cite{Brey2006,Akhmerov2008} However, in our case, no atomically
defined boundary exists between the central region and the bordering GAL regions and we thus adopt a
model wherein the band gap, $E_g$, of the confining GALs is included via a position-dependent mass term. 
We introduce dimensionless 
coordinates 
$(\chi,\gamma)=2/W \times (x,y)$
and dimensionless energies $\epsilon= E/E_0$, with 
$E_0=2\hbar v_F/W\simeq 1.278$~eV$\cdot$nm$\times W^{-1}$, assuming a Fermi velocity
$v_F=10^6$~m/s of graphene. 
In these units, the eigenvalue problem for the spinor eigenstates reads
\begin{equation}
\left[\!\!
\begin{array}{cc}
m(\chi) & -i\left(\partial_\chi - i\partial_\gamma\right) \\
-i \left(\partial_\chi + i\partial_\gamma\right) & -m(\chi)
\end{array}
\!\!\right]
\!\!
\left[\!\!
\begin{array}{c}
\phi_A(\chi,\gamma) \\ \phi_B(\chi,\gamma)
\end{array}
\!\!\right]
=\epsilon
\left[\!\!
\begin{array}{c}
\phi_A(\chi,\gamma) \\ \phi_B(\chi,\gamma)
\end{array}
\!\!\right],
\end{equation}
where the dimensionless mass term $m(\chi)=\delta=E_g/(2E_0)$ for $|\chi|>1$ and zero otherwise.
This equation has been derived from the Dirac Hamiltonian, $H_K$, which emerges as a linearization of a
TB model of graphene near the $K$ point. We will discuss the differences between the two inequivalent
$K$ and $K^\prime$ valleys of graphene below.
Note that the sign of the mass term is arbitrary, and that changing it has no physical significance, but merely
results in an interchange of the two spinor components. We stress, however, that the sign should be the
same on both sides of the central, waveguiding region.
The Hamiltonian commutes with the $y$--component of the momentum operator and we
thus take spinor components of the form $\phi_A(\chi,\gamma)=f(\chi)e^{i\kappa\gamma}$ and 
$\phi_B(\chi,\gamma)=g(\chi)e^{i\kappa\gamma}$, where $\kappa=kW/2$ is the 
dimensionless Bloch wave vector along the longitudinal direction of
the waveguide. As the mass term is piecewise constant, the equations for the spinor components can
be decoupled in each region of the waveguide structure. 
We look for bound states, and thus take $\epsilon^2<\delta^2$. The normalizable solutions for the first
spinor component thus read 
$f(\chi)=A_\pm e^{\pm\beta \chi}$ for $|\chi|>1$ and 
$f(\chi)=B\cos(\alpha \chi)+C\sin(\alpha \chi)$
for $|\chi|<1$ while the second component is given via 
$g(\chi) = i(\kappa-\partial_\chi)f(\chi)/(\epsilon+m(\chi))$. 
Here, we have defined $\alpha=\sqrt{\epsilon^2-\kappa^2}$ and $\beta=\sqrt{\delta^2+\kappa^2-\epsilon^2}$.
The requirement of continuity of both spinor components at the boundaries of the central waveguide region
leads to a transcendental equation for the energies, $\sqrt{\delta^2-\alpha^2} = \alpha\tan{2\alpha}$,
regardless of which valley is considered (see below).
The solution of the problem is thus closely reminiscent of that of an ordinary one-dimensional square 
well potential, albeit with the crucial difference that for graphene $E\propto \alpha$ rather than 
$E\propto \alpha^2$, as a consequence of the linear dispersion relation of graphene.
Also, we note that, contrary to the case of the Schr\"odinger equation, the derivative of the 
eigenstate spinors need not be continuous at the boundary. 

\subsection{Infinite mass limit}
The standard textbook graphical solution suggests that there are $N$ bound states in the wave\-guide
if $(N-1)\pi/2 < \delta < N\pi/2$. In general, the energies and spinor components of these states will need
to be determined by numerical solution of the transcendental equation. However, in the limit of an
infinite mass term, $\delta\rightarrow \infty$, the problem admits an analytical solution for the energies,
\begin{equation}
\epsilon_{ns}^{\infty}(\kappa) = s\sqrt{\kappa^2+\frac{\pi^2}{4}\left(n+\tfrac{1}{2}\right)^2},\;n=0,1,2\ldots
\end{equation}
or, reverting to ordinary units,
\begin{equation}
E_{ns}^{\infty}(k) = s\sqrt{E^2_b(k) + \frac{\hbar^2 v_F^2\pi^2}{W^2}\left(n+\tfrac{1}{2}\right)^2},
\end{equation}
where $E_b(k) = \hbar v_F k$ is the bulk graphene dispersion relation and $s=\pm 1$. 
The dispersion relations of the waveguide modes thus resemble those of gapped graphene\cite{Pedersen2009} with a
mass term of $\Delta_\mathrm{eff}=(\hbar v_F \pi/W) (n+\frac{1}{2})$.
Series expansion of the transcendental equation for the energies in the case of a finite mass term
reveals that $\alpha\simeq \pi/2(n+1/2)[1+1/(2\delta)]^{-1}$ so the results in the infinite mass limit
are expected to be valid when $E_g\gg 2\hbar v_F/W\simeq 1.278$~eV$\cdot$nm$\times W^{-1}$.
Including the leading order correction, the energies are given as
\begin{equation}
E_{ns}^{(1)}(k) = E_{ns}^{\infty}(k) -s\frac{2\hbar^2v_F^2\pi^2(n+\tfrac{1}{2})^2}{W^2\sqrt{k^2W^2+\pi^2(n+\tfrac{1}{2})^2}}\times\frac{1}{E_g}.
\end{equation}

The eigenstate spinors in the infinite mass limit are most easily determined by using the boundary conditions
derived by Berry,\cite{Berry1987} which set a phase relationship between the spinor 
components at the edge of the waveguide.
The spinor components for $\kappa=0$ and $\epsilon>0$ can then be derived as
\begin{eqnarray}
f_n(\chi) &=&
\left\{\begin{array}{ll}
\cos\left(\pi/2[n+\tfrac{1}{2}] \chi \right) & \mathrm{for~}n\mathrm{~zero~or~even}, \\
\sin\left(\pi/2[n+\tfrac{1}{2}] \chi \right) & \mathrm{for~}n\mathrm{~odd},
\end{array}\right.,
\end{eqnarray}
while
\begin{eqnarray}
g_n(\chi) &=&
\left\{\begin{array}{ll}
i\sin\left(\pi/2[n+\tfrac{1}{2}] \chi \right) & \mathrm{for~}n\mathrm{~zero~or~even}, \\
-i\cos\left(\pi/2[n+\tfrac{1}{2}] \chi \right) & \mathrm{for~}n\mathrm{~odd},
\end{array}\right.,
\end{eqnarray}
for $|\chi|<1$.
Here, we have omitted normalization constants. We note that because of the particle-hole symmetry of graphene, 
the eigenstates should be normalized separately on each sublattice.\cite{Brey2006a}
The spinors for the $\epsilon<0$ eigenstates are given by exchanging $f_n$ and $g_n$. 
Interestingly, and in stark contrast to an ordinary infinite square well potential, this shows 
that in the limit of infinite mass, the charge density is evenly distributed within the waveguide, for states
with vanishing wave vectors.
We find that this holds true also in the more
general case of a finite mass term, the main difference being that in this case the spinors extend slightly
into the mass regions. Including a nonzero wave vector, the charge density is localized predominantly
on either the edges or the middle of the waveguide.

The derivation above takes as its starting point a linearization of graphene near the $K$ valley.
The Dirac Hamiltonian near the inequivalent $K^\prime$ valley is such that the energy spectrum of the waveguide is the same
for both valleys, while the eigenstate spinors are related by an interchange of the spinor components, i.e.
$\phi_A^{K^\prime}=\phi_B^K$ and $\phi_B^{K^\prime}=\phi_A^K$. 
We also note that to obtain the full
wave function, a Bloch phase factor of $e^{i\mathbf{K}\cdot\mathbf{r}}$ or $e^{i\mathbf{K}^\prime\cdot\mathbf{r}}$ 
should be added to the spinors, with $\mathbf{K}$ and $\mathbf{K}^\prime$ depending on the orientation of the graphene lattice. 
We will return to this point below, when
we compare the results of the Dirac approximation with those obtained with a tight-binding model.

\section{Tight-binding model}
To include the atomistic details of the waveguide structures, 
and to clarify the validity of the results derived in the Dirac approximation,
we employ a nearest-neighbor tight-binding approximation.
This will also allow us to asses the significance of the orientation of the waveguide with respect to the 
graphene lattice. The TB model is parametrized via a hopping term $t=-3$~eV between $\pi$--orbitals, 
the on-site energy of which we set to zero. We ignore non-orthogonality of the $\pi$--orbitals.
Fig.~\ref{fig:concept} illustrates the geometry used for the TB model in the case of a $\{7,3\}_2^\mathrm{zz}$
GAL waveguide. We use periodic boundary conditions along the $x$--axis as well as the $y$--axis. 
We have ensured that the results are converged with respect to the number of GAL unit cells
included around the waveguide. The band gap of the GAL is quite well developed even with just a
few rows of antidots,\cite{Gunst2001} so including three GAL unit cells on each side of the
waveguide usually yields converged results.

\section{Results}
\subsection{Dispersion relations}
\begin{figure}
\includegraphics[width=\linewidth]{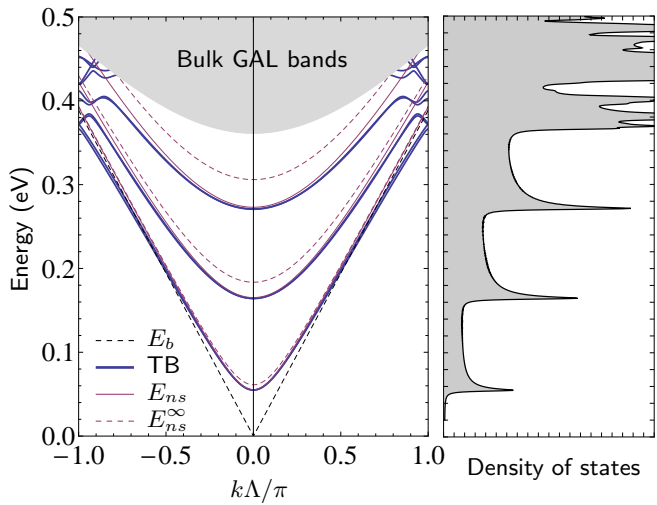}
\caption{(left) Band structure of the $\{7,3\}_5^\mathrm{zz}$ GAL waveguide.
The band structures are shown for the tight-binding model as well as the analytical
infinite mass-limit results, $E_{ns}^\infty$ and numerical solution of the transcendental
equation of the Dirac approximation, $E_{ns}$. The shaded gray area shows the projected
bands of the surrounding GAL regions. 
For comparison, the bulk graphene band structure, $E_b$, is also shown. 
Note that $\Lambda$ denotes the lattice constant of the waveguide.
(right) Corresponding density of states for the TB model.
Note the van Hove singularities characteristic of one-dimensional structures.
}
\label{fig:bandL7R3W5t}
\end{figure}
In Fig.~\ref{fig:bandL7R3W5t}, we show the band structure of a $\{7,3\}_5^\mathrm{zz}$ GAL waveguide calculated
using the TB model as well as the Dirac approximation. 
Only electron ($E>0$) bands are shown. Both the TB and the Dirac model exhibit perfect electron-hole
symmetry, so hole ($E<0$) bands simply follow from a sign change.
For the Dirac results we take the effective
width of the waveguide to be $W_\mathrm{eff}=(N+\tfrac{1}{2})w$, slightly larger than the definition used for
denoting the waveguide geometries. 
Note that the wavevector is shown relative to the lattice constant of the GAL waveguide, which is
$\Lambda=3La$ for the zigzag orientation and $\Lambda=\sqrt{3}La$ for the armchair orientation.
In the Figure, the shaded, gray region illustrates the projected bands of the GAL, which define
the region below which localized waveguide states are expected to appear. 
This particular waveguide structure supports several localized states. 
Higher-lying band gaps also appear in the GAL, and we have confirmed that localized waveguide modes are 
also supported in these gaps. 
It is worth stressing that localized waveguide modes exist for all wave vectors in the first Brillouin zone. 
This is in contrast to the case of waveguides defined via electrostatic gating, where guided modes
generally exist only for a limited range of wave vectors.\cite{Zhang2009}
The dispersions of the waveguide states agree very well between the
TB and the Dirac model, as long as the wave vector is not too near the Brillouin zone edges.
The largest deviations between the two models occur for energies close to the projected bands of
the GAL, where coupling between the waveguide and the GAL states is pronounced.
We note that the analytical result obtained in the infinite mass limit, $E_{ns}^\infty$ is a very 
good approximation of the lowest waveguide mode. Including the first-order correction to $E_{ns}^\infty$ 
leads to near-perfect agreement with the full solution of the transcendental equation.
Note that as illustrated in the derivation of the Dirac result, the waveguide dispersion relation quite 
closely resembles that of gapped graphene, which in turn is approximately the same as bulk graphene 
for wavevectors $k\gg \Delta_\mathrm{eff}/(\hbar v_F)$. This is evident in the figure, where for 
comparison we also include the bulk graphene dispersion relation.
The density of states (DOS) calculated using the TB model is shown in the right panel of
Fig.~\ref{fig:bandL7R3W5t}. As expected from the Dirac approximation, the DOS closely resembles 
one-dimensional gapped graphene, i.e. 
$g(E)\propto \Theta(E-\Delta_\mathrm{eff})E/(\hbar v_F  \sqrt{E^2-\Delta_\mathrm{eff}^2})$ with
the van-Hove singularities characteristic of one-dimensionality clearly evident in the figure.

\subsection{Band gaps and effective masses}
\begin{figure}
\includegraphics[width=.9\linewidth]{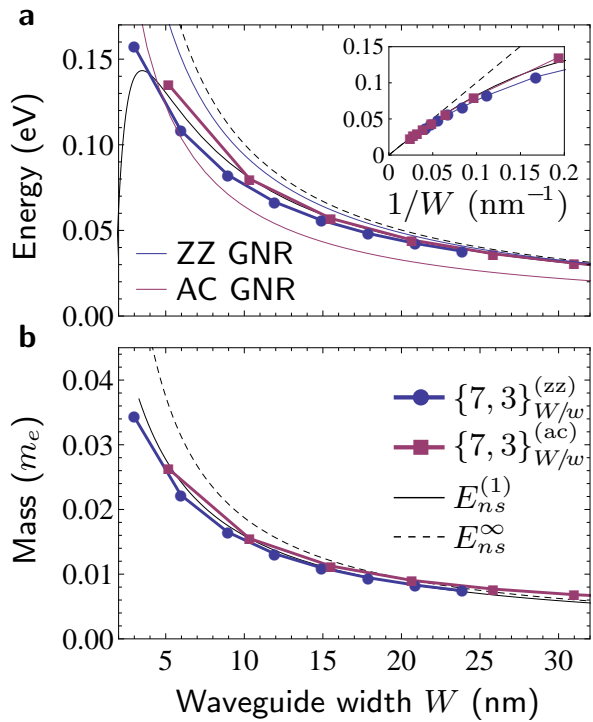}
\caption{(a) Energy of the lowest localized waveguide state at the $\Gamma$ point, for the $\{7,3\}_N$ family of GAL waveguides.
The energy is shown as a function of the width of the waveguide.
Results are shown for TB models (points) of waveguides oriented along the zigzag (ZZ) and armchair (AC) directions, respectively, as
well as the analytical results obtained via the Dirac model in the infinite mass limit and including the first-order correction (black lines). See the legend in panel b.
The inset illustrates the $1/W$ dependence of the energy for wide waveguides.
For comparison, the thin colored lines show the energies for ZZ and AC GNRs, if certain edge dependencies are 
ignored (see text).
(b) Corresponding effective masses in units of the free electron mass, $m_e$. In both panels, note the close resemblance
of the results obtained for AC and ZZ oriented waveguides.
}
\label{fig:mEvsW}
\end{figure}
In Fig.~\ref{fig:mEvsW}a, we show the energy of the lowest localized waveguide state at the $\Gamma$ point as a function
of the width of a $\{7,3\}_N^\mathrm{(zz,ac)}$ waveguide. Results are shown for the TB model as well as the analytical results obtained
in the Dirac equation approach. 
Note that due to electron-hole symmetry, the band gap is twice this value.
While the solution derived from the Dirac equation does not distinguish between zigzag 
and armchair orientation of the waveguide, the TB model predicts that there are some differences between the two cases.
To illustrate this, we show results obtained for waveguides oriented along the zigzag (ZZ) and armchair (AC) directions, respectively.
The inset of the figure illustrates that both ZZ and AC orientations exhibit a clear $1/W$ dependence of the energies, as predicted 
from the Dirac approximation, provided the waveguide is sufficiently wide.
We note that while differences do exist between AC and ZZ orientations, these are rather small, and could be attributed to
a slightly different effective width of the waveguides in the two cases, due to the $\pi/6$ difference in the orientation of the surrounding 
GAL with respect to the waveguide. Indeed, results of the Dirac approximation fit both orientations quite well, especially when
including the first-order correction. The results of the Dirac model can be made to fit even better if we take into account the fact
that the effective width of the waveguide is likely to be somewhat larger than the definition we have used, see Fig.~\ref{fig:concept}. 
Indeed, introducing the same effective width in the Dirac model as we did for Fig.~\ref{fig:bandL7R3W5t} results in even better agreement 
with TB results.

The absence of a well-defined edge means that the dependence of the properties of the guided modes
on the width of the waveguide is much simpler than is the case for GNRs. In particular, 
GAL waveguides are always semiconducting, whereas in a nearest-neighbor TB model armchair GNRs alternate 
between metallic and semiconducting behavior depending on the exact width of the ribbon, while zigzag GNRs display
dispersionless midgap states, localized on the edges.\cite{Brey2006} In Fig.~\ref{fig:mEvsW}a, we show the energies 
for ZZ and AC GNRs, calculated via the TB model. We stress that the ZZ GNRs also contain dispersionless edge
states at the Dirac point energy. Furthermore, AC GNRs are metallic for widths $W=(3p-1)a$, with $p$ an integer.\cite{Nakada1996} To compare with
the waveguide results, we have included only semiconducting AC GNRs in the figure. With these modifications, there is quite
good agreement between the energies of GNRs and the GAL waveguide structures. GAL waveguides thus resemble ribbons without
the particulars resulting from edge effects. As such, we speculate that the electronic properties of GAL waveguides may be 
easier to control experimentally.

In Fig.~\ref{fig:mEvsW}b, we show the effective mass of the lowest waveguide state at the $\Gamma$ point as a function
of the width. From the analytical Dirac results in the limit of an infinite mass term, we find an effective mass
$m_\mathrm{eff}^{\infty}=\hbar\pi(n+\tfrac{1}{2})/(v_F W)$. Again, we note that there is excellent agreement between
the results obtained in the Dirac approximation and those from a TB model, for both orientations of the waveguide.
Including the first-order correction, results from the Dirac model are in near-perfect agreement with the TB model.
While electrons in pristine graphene have vanishing effective masses, the appearance of an effective mass term in the
waveguide structures results in non-zero, albeit still very small effective electron masses, which tend to zero in the
limit of infinitely wide waveguides.

\subsection{Eigenstates}
\begin{figure*}
\includegraphics[width=.9\linewidth]{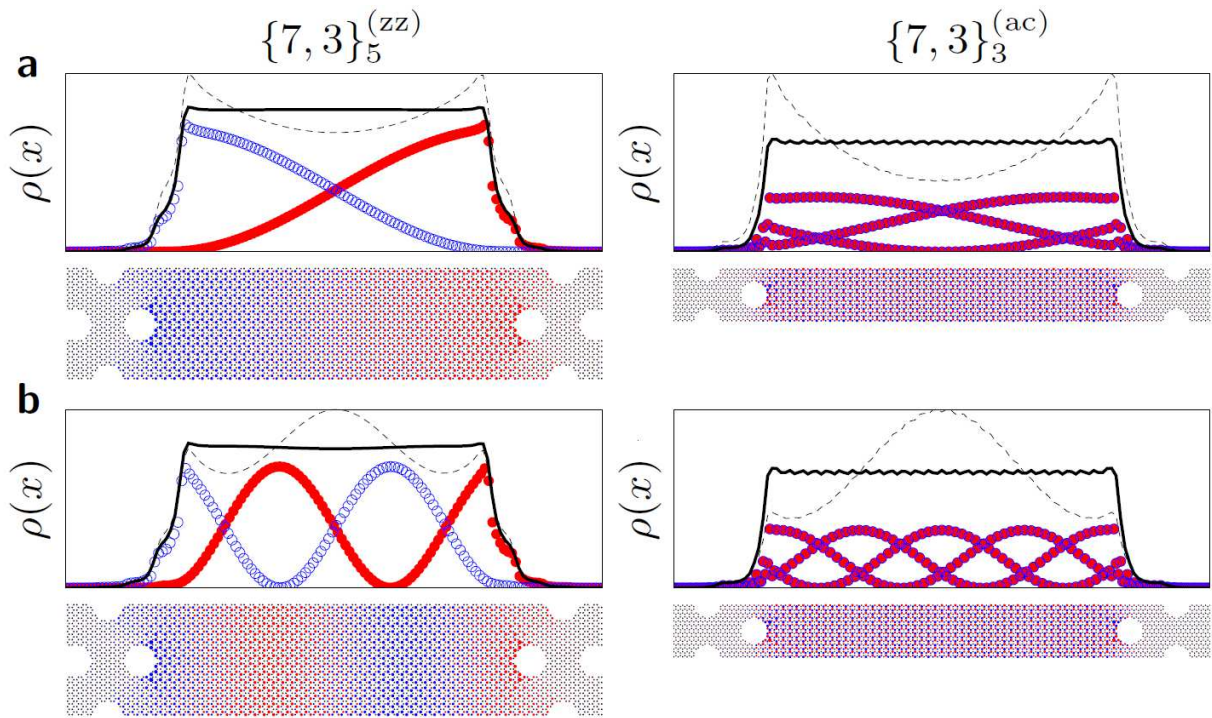}
\caption{
Wave functions of the localized waveguide modes corresponding to the (a) lowest and the (b) second-lowest (positive) 
energy at the $\Gamma$ point of a (left) $\{7,3\}_5^\mathrm{(zz)}$ and a (right) $\{7,3\}_3^\mathrm{(ac)}$ GAL waveguide. 
The lower panels in each case show
the geometry, with carbon atoms indicated with black dots. Note that the actual computational cell includes additional GAL
unit cells on each side of the central region.
Superimposed on top of the geometry is the wave function, 
with the size of the circles indicating the absolute square of the $\pi$--orbital coefficient, while the color
indicates sublattice. The upper panels show the integrated probability density, $\rho(x)\equiv \int|\psi(x,y)|^2 dy$.
Red and blue circles indicate the densities on each sublattice. The black line shows the total density, when including
a small broadening term. The dashed line shows the corresponding density at non-zero wavevectors, $k\Lambda=\pi/2$.
Note the rapid oscillations of the integrated density of the AC waveguide.
}
\label{fig:states}
\end{figure*}
To further compare the AC and ZZ waveguide orientations, we show in Fig.~\ref{fig:states}a 
the eigenstates corresponding to the lowest (positive) energy of $\{7,3\}_5^\mathrm{(zz)}$ and $\{7,3\}_3^\mathrm{(ac)}$ 
GAL waveguides, calculated at the $\Gamma$ point. Note that these have approximately the same waveguide widths.
The lower panels in the figure show the absolute square of the wave function, with the color indicating the sublattice.
These results demonstrate a crucial difference between the AC and ZZ orientations, namely that while for the AC waveguide,
the wave function is distributed evenly across the two sublattices, the ZZ waveguide exhibits pseudo-spin polarization, with
the wave functions of the two sublattices localized predominantly on opposite edges of the waveguide. We note that the
lowest energy is doubly degenerate, and that the second eigenstate (not shown) has the opposite pseudo-spin distribution.
The upper panels in the figure show the integrated probability density, $\rho(x)\equiv \int|\psi(x,y)|^2 dy$ along the
transversal direction of the waveguide, with color indicating the sublattice. Note that despite the lack of a clearly defined
edge, the probability densities very closely resembles those of GNRs.\cite{Brey2006} The black line in the upper panels show
the total probability density, if a broadening of the order of the graphene lattice constant is included in order to account
for the spatial extent of the $\pi$--orbitals. As predicted from the Dirac equation approach above, these results illustrate
how the charge density is nearly uniformly distributed across the entire waveguide, also for the higher-lying states shown
in Fig.~\ref{fig:states}b. This is in stark contrast to gate defined waveguides, which have wave functions more reminiscent 
of ordinary standing wave solutions.\cite{Zhang2009} In agreement with the Dirac results, we find that the uniform distribution
only occurs at the $\Gamma$ point. For non-zero wavevectors the density becomes localized predominantly at the edges of
the waveguide for the lowest state, as illustrated with dashed lines in the figure for $k\Lambda=\pi/2$. In contrast to this, 
the densities of the second-lowest states tend to localize in the center of the waveguide as the wavevector is increased.

To compare the wave functions with the spinor components derived via the Dirac equation, we first note that both the $K$ and the
$K^\prime$ points of graphene are folded onto the $\Gamma$ point of the waveguide structure. We thus expect the eigenstates
to resemble linear combinations of the eigenstates in both valleys. The differences between the zigzag and armchair waveguides 
emerge due to the Bloch phase factors $e^{i\mathbf{K}\cdot \mathbf{r}}$ and $e^{i\mathbf{K^\prime}\cdot \mathbf{r}}$, which differ 
depending on the orientation of the graphene lattice. For the zigzag orientation, the integrated probability density, $\rho(x)$, becomes a 
simple linear combination of the eigenstates belonging to each valley, 
$\rho_A(x)\propto |f_K(x)+f_{K^\prime}(x)|^2$ 
and 
$\rho_B(x)\propto |g_K(x)+g_{K^\prime}(x)|^2$. Using the expressions for the spinor components derived above, we
find $\rho(x)\propto 1\pm\sin(2\pi[n+\tfrac{1}{2}]x/W)$, with the sign depending on the sublattice, which is in
excellent agreement with the TB results.
In contrast to this, because of the mixing of the valleys the probability densities of the armchair orientation exhibit
a rapidly oscillating term, with a period $2\pi/\Delta K$.\cite{Brey2006a} 
These rapid oscillations are clearly evident in the right panels of Fig.~\ref{fig:states}. As shown in the figure, these
rapid oscillations are quickly washed out if a small amount of broadening is included. In this case, we recover the nearly uniform
charge distribution within the waveguide predicted from the Dirac results.
Finally, we note that the differences between armchair and zigzag oriented GAL waveguides are very similar to those seen 
between GNRs with armchair and zigzag edges.\cite{Brey2006} In the case of GNRs, though,
the difference emerges due to different boundary conditions at the edge of the ribbon, which are not present in the
case of GAL waveguides.

\begin{figure}
\includegraphics[width=.9\linewidth]{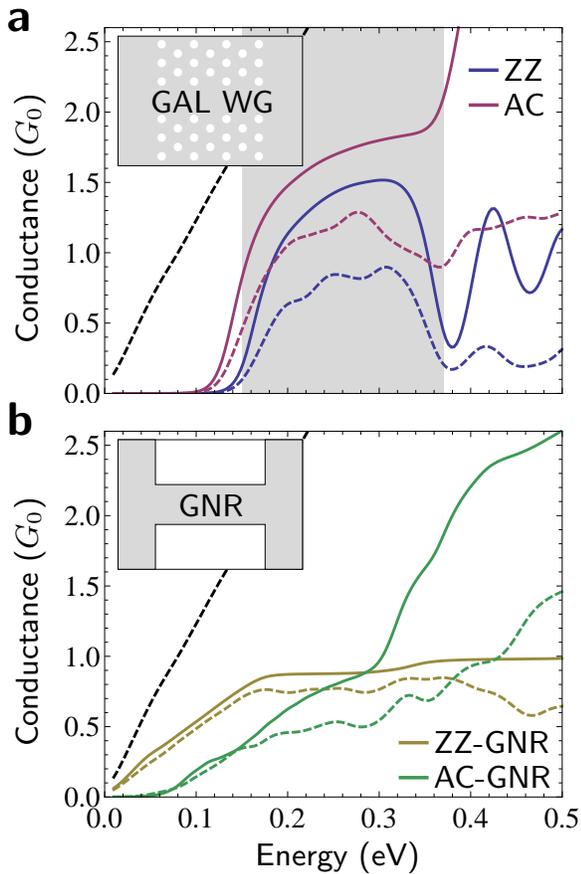}
\caption{Insets show the schematic transport setups for (a) a GAL waveguide and (b) a GNR, each connecting two semi-infinite graphene leads. Conductances for pristine (solid lines) and disordered systems (dashed lines) are shown for (a) GAL waveguides and (b) GNRs, in units of the conductance quantum $G_0$. The straight black dashed lines show the graphene conductance. In the disordered systems, edge atoms have been randomly removed with a 5\% probability. The length of the waveguides and GNRs are $L=89\,$nm. The shaded area in (a) indicates the energy range of the confined waveguide mode.}
\label{fig:G}
\end{figure}

\subsection{Conductance}
Because the GAL waveguides have no clearly defined edge, one might wonder whether the guiding properties of the waveguides would be relatively robust to disorder. Indeed, the crucial ingredient is the existence of a band gap in the surrounding GAL. As this gap essentially occurs due to an averaging of the effect of the individual holes,\cite{Pedersen2008a} the emergence  of a gap should be relatively robust to disorder. 
A thorough investigation of disorder is beyond the scope of this paper, but as a preliminary study we model disorder by randomly removing atoms at the edges of the holes in the GAL.  We consider a disordered GAL waveguide sandwiched between semi-infinite pristine graphene leads as illustrated in the inset of Fig. \ref{fig:G}a. For comparison we also consider the analogous system with the two graphene leads connected with a GNR having the same width 
as the corresponding waveguide ($W=4.5\,$nm for ZZ, $W=6.0\,$nm for AC), as illustrated in the
inset of Fig.~\ref{fig:G}b. We calculate the transmission through the waveguide and GNR using a recursive Green's function method\cite{MarkussenPRB2006} with the lead-self energies determined using an iterative procedure.\cite{Sacho1984} The transmission is averaged over 100 values of the transverse wave vector, and we further average over 10 samples with different realizations of the random disorder. To smear out the rapid oscillations that occur due to interference between transmitted  and reflected waves at the boundaries between the GAL waveguide and the graphene leads, we calculate the conductance at a finite temperature of $T=100$~K. We consider a relatively high level of disorder, for which edge atoms are removed with a 5\% probability. In Fig.~\ref{fig:G}a, we show the conductance of disordered $\{7,3\}_1^\mathrm{(zz,ac)}$ GAL waveguides of length $L=89\,$nm (dashed lines).  For comparison, we also include the conductance of the pristine waveguides (solid lines). The shaded area indicates the energy range for the confined waveguide mode. In Fig.~\ref{fig:G}b, we show the corresponding results for ZZ- and AC-GNRs. 

Focusing first on the conductances for the pristine systems (solid lines), we observe that the ZZ- and AC oriented GAL waveguides have a similar conductance in the energy range of the waveguide mode. The high conductances $G\sim1.5\,G_0$ show that there is a relatively good electronic coupling between the GAL waveguide and the graphene leads. The metallic ZZ-GNR is conducting at all energies, but the transmission saturates at a value close to unity. In the energy range of the waveguide mode, the GAL waveguides thus have a higher conductance than both of the GNRs. 
Turning to the results for the disordered systems we observe that the GAL waveguides retain relatively high conductances. The ZZ and AC orientations show comparable reductions in the conductance due to disorder. The ZZ-GNR is less affected by disorder, while the AC-GNR conductance is significantly reduced except at the lowest energies.
We conclude that the GAL waveguides appear to be relatively robust against structural disorder and in general have higher conductances than the corresponding GNR systems.

\subsection{Waveguide bends}
\begin{figure}
\includegraphics[width=\linewidth]{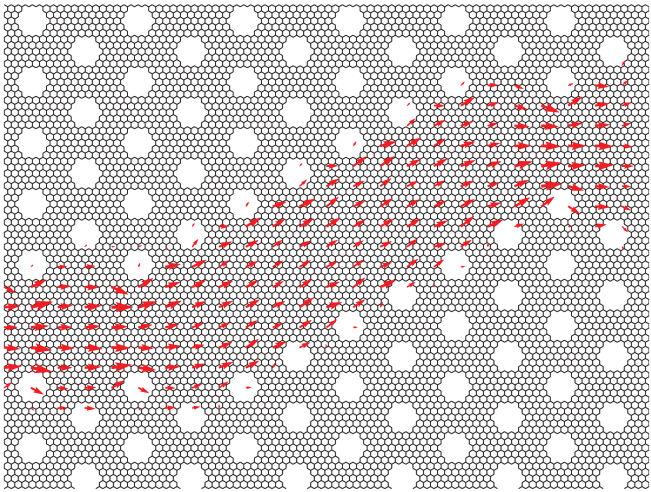}
\caption{
Bond current through a 'kinked' $\{5,2\}_1^\mathrm{(zz)}$ GAL waveguide. The current is calculated at energy $E=0.25$eV with a transmission of $\mathcal{T}=1.9$. The current is highly confined to the waveguide region and no additional reflections are observed due to the kinks. }
\label{fig:kinkBC}
\end{figure}
As mentioned earlier, the waveguides introduced in this paper are closely analogous to photonic crystal waveguides. In such
structures, light can be guided through bends in the waveguide with very little radiation loss.\cite{Chutinan2000}
Relying on this analogy, we expect GAL waveguides to show a similar robustness to kinks along the waveguide.
To illustrate the localization of the electronic state and the guiding properties of the GAL waveguide we show the local current through the waveguide in Fig.~\ref{fig:kinkBC}. Similarly to the results of Fig.~\ref{fig:G}, the waveguide structure is connected on both sides to semi-infinite bulk graphene leads, not included in the figure. The left going bond current (per unit energy) in the presence of an infinitesimal bias voltage is calculated from the left scattering state spectral function, $A_{L,ij}$, and the hopping matrix elements $H_{ij}$ from the TB Hamiltonian. Between atom $i$ and $j$ the local current is $A_{L,ij}\, H_{ij}$.\cite{Todorov2002} To visualize the current on the given scale the current running away from each atom was calculated and averaged over an applied mesh. The illustrated average current thus cannot be assigned to the individual atoms anymore, which is the reason why current appears to occur within the holes in the figure.
Since the $k$-averaged transmission only changes slightly from the $\Gamma$-point result, we use the $\Gamma$-point scattering states. Fig.~\ref{fig:kinkBC} clearly illustrates the confinement of the current to the waveguide region and the robustness against kinks. 

\begin{figure}
\includegraphics[width=.9\linewidth]{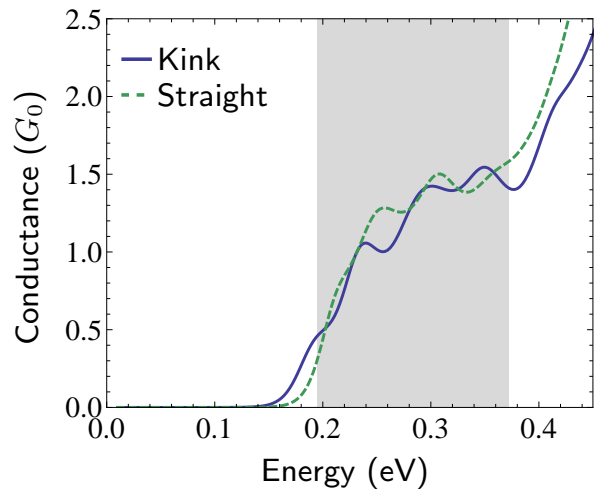}
\caption{Conductances of the 'kinked' waveguide shown in Fig.~\ref{fig:kinkBC} (straight line) and of a straight waveguide of similar length (dashed line). The shaded area indicates the energy range of the confined waveguide mode of the straight waveguide. Note the nearly identical
conductances of the two systems.}
\label{fig:Gkink}
\end{figure}
To further illustrate the strong guiding properties of GAL waveguides, we show in Fig.~\ref{fig:Gkink} the conductance of the waveguide bend
illustrated in Fig.~\ref{fig:kinkBC}. A similar method was used as that for the results of Fig.~\ref{fig:G}. For comparison, we also show the corresponding
conductance through a waveguide generated by omitting the kink in Fig.~\ref{fig:kinkBC} and instead having the waveguide run
straight through. These results show that while there are small differences between the two structures
in the oscillations of the conductance, overall the introduction of a kink has almost no consequence on the conductance through
the GAL waveguide. Very low reflection loss is thus introduced by the kink, despite the fact that the waveguide alternates between 
zigzag and armchair orientations. We note that we have found similar results for slightly different waveguide structures.

\section{Discussion and summary}
The GAL waveguide systems studied in this paper represent idealized structures, which may be challenging to realize experimentally due to the small hole sizes. However, the applicability of the Dirac model allows for simulations of arbitrarily large structures. Moreover, the Dirac model can equally well be applied to other gapped graphene systems, where the band gap is not induced through periodic holes, but e.g. via periodically absorbed hydrogen.\cite{Balog2010}
Although the Dirac and TB models applied in this work are very simple, previous studies of pure GALs have shown that both the Dirac- and TB models reproduce the trends obtained from more accurate density functional theory calculations.\cite{Furst2009} In any case, the concept of a GAL waveguide depends only on the existence of a band gap in the GAL region and not on the specific details, and we believe our simplified models capture the correct physics.

In summary, we have introduced GAL waveguides. The band structures of GAL waveguides have been modeled with the Dirac model including a mass term, which is shown to be in excellent agreement with an atomistic tight-binding model. We have shown that GAL waveguides support modes which are highly confined to the waveguide region and are robust against structural disorder and kinks in the waveguide. 
In transport calculations, we find that GAL waveguides have higher conductances than corresponding graphene nanoribbons. A further advantage of the surrounding GAL may be that it will mechanically stabilize the structure and be able to carry some of the generated Joule heat away from the device. GAL waveguides may thus be an attractable way of realizing electronic wires in integrated graphene circuits.

\section{acknowledgement}
The work by J.G.P. is financially supported by the Danish
Council for Independent Research, FTP Grants No. 11-105204
and No. 11-120941. T.M. acknowledges support from  the Danish
Council for Independent Research, FTP Grants No. 11-104592 and No. 11-120938. The Center for Nanostructured Graphene
(CNG) is sponsored by the Danish National Research Foundation.


\end{document}